# Robust and discriminative zero-watermark scheme based on invariant feature and similarity-based retrieval for protecting large-scale DIBR 3D videos

Xiyao Liu[1], Yifang Wang[1], Ziqiang Sun[2]*, Lei Wang[1], Rongchang Zhao[1], Yuesheng Zhu[3] and Beiji Zou[1]

1. School of Computer Science and Engineering, Central South University, Changsha, 410083, China.
2. Nanjing Research Institute of Electronics Technology, Nanjing, 210039, China.
3. Shenzhen Graduate School, Peking University, Shenzhen, 518055, China.

E-mail: lxyzoewx@csu.edu.cn; sunziq87@163.com    Telephone:+8615111327037; +8615951832106

**Abstract**: Digital rights management (DRM) of depth-image-based rendering (DIBR) 3D video is an emerging area of research. Existing schemes for DIBR 3D video cause video distortions, are vulnerable to severe signal and geometric attacks, cannot protect 2D frame and depth map independently or can hardly deal with large-scale videos. To address these issues, a novel zero-watermark scheme based on invariant feature and similarity-based retrieval for protecting DIBR 3D video (RZW-SR$_{3D}$) is proposed in this study. In RZW-SR$_{3D}$, invariant features are extracted to generate master and ownership shares for providing distortion-free, robust and discriminative copyright identification under various attacks. Different from traditional zero-watermark schemes, features and ownership shares are stored correlatively, and a similarity-based retrieval phase is designed to provide effective solutions for large-scale videos. In addition, flexible mechanisms based on attention-based fusion are designed to protect 2D frame and depth map independently and simultaneously. Experimental results demonstrate that RZW-SR$_{3D}$ have superior DRM performances than existing schemes. First, RZW-SR$_{3D}$ can extracted the ownership shares relevant to a particular 3D video precisely and reliably for effective copyright identification of large-scale videos. Second, RZW-SR$_{3D}$ ensures lossless, precise, reliable and flexible copyright identification for 2D frame and depth map of 3D videos.

**Keywords**: DIBR 3D video; zero-watermark scheme; invariant feature; similarity-based retrieval; attention-based fusion; copyright identification

# 1. Introduction

Protecting the copyrights of three-dimensional (3D) videos has become a crucial issue [4, 22, 29 ,41]. On one hand, the risk of copyright infringement for 3D videos has been increasing because 3D videos, which can provide better immersive experiences to viewers than traditional 2D videos, have been becoming more and more popular over the Internet [18, 28, 36, 42]. One the other hand, the illegal copying and distribution of 3D videos will cause more serious losses to their owners than of 2D frames because the production costs of 3D videos are much higher.

In this study, DIBR 3D video is chosen as the protection target due to the following two reasons. First, 3D videos stored in DIBR format can save storage and transmission-bandwidth compared with stereoscopic format because the depth maps, which only consist of gray-level pixels with smoothed areas, are amenable to effective compression [27, 36]. Second, existing 2D videos can be converted to 3D videos with their calibrated depth maps based on DIBR technique [5, 23, 44], and 3D videos synthesized in this manner cost much lower because they do not need different cameras to capture video frames from various viewpoints.

Compare with protecting traditional 2D videos, protecting DIBR 3D videos is more complex. The original 2D frame with depth map in 3D videos can converted into synthesized frames. Therefore, the watermark should be obtained from any one of original 2D frame, synthesized frames or depth map. Because synthesized frames after DIBR are different from original 2D frame and their pixels are shifted horizontally from original one, it means that the protecting scheme for 3D videos should be DIBR invariant. In addition, the generators of 2D frame and depth map are possibly different for 3D videos synthesized from existing 2D videos. In this situation, copyrights of 2D frame and depth map should be protected independently. Otherwise, they should be protected simultaneously.

Watermark is a one of most popular methodologies to tackle the digital rights management (DRM) issue [4, 21, 35, 40] but existing watermark schemes for DRM of DIBR-based 3D videos, which can be mainly classified into 2D frame-based watermark, depth map-based watermark and zero-watermark schemes, still have rooms for improvements. 1) For 2D frame-based watermark schemes [1-3, 6-7, 9, 13, 16, 19, 22, 27, 30, 32, 43] which only embed watermark into the 2D frame, they introduce irreversible distortions to the video content [29]. In addition, they ignore the situation that generators of 2D frame and depth map may be different and cannot protect the copyright of depth map independently. 2) For depth map-based watermark schemes [8, 24-25, 29, 33-34, 38] which only embedded watermark into depth map, their robustness against severe signal attacks and geometrical attacks is insufficient [26]. In addition, they cannot protect the copyright of 2D frame independently. Finally, they cannot extract any copyright information from the 3D videos transmitted and stored by two synthesized frames. 3) For zero-watermark scheme [26], it generates ownership shares, which indicates the mapping ships between video feature and watermark, without any direct watermark embedding. Its robustness against geometrical attacks still needs improvements. More importantly, it is based on the premise that the precise ownership share relevant to a particular video has been already obtained for effective copyright identification. Unfortunately, it is difficult for zero-watermark to obtain the precise ownership share effectively when processing large-scale 3D videos due to the lack of efficient indexing mechanism.

To address the above-mentioned issues, we proposed a novel zero-watermark scheme for copyright protection of DIBR 3D video (RZW-SR$_{3D}$) by fusing the similarity-based retrieval into the architecture of traditional zero-watermark in this study. To the best of our knowledge, it is the first scheme that fuses similarity-based retrieval with zero-watermark techniques for protecting large-scale DIBR 3D videos. Zero-watermark scheme is chosen because it is distortion-free, more robust and can protect the copyrights of 2D frame and depth map simultaneously and independently, which outperforms the 2D frame-based and depth-based watermark schemes. Moreover, a similarity-based retrieval mechanism is designed and fused into the architecture of traditional zero-watermark to deal with large-scale DIBR 3D videos.

We design robust and discriminative features by calculating the centroids of normalized deviations between temporally informative representative images [11] and video frames. The features extracted in this manner are invariant under various signal, geometric and DIBR distortions and thus can be utilized for effective similarity-based retrieval and copyright identification. In our proposed scheme, features of 2D frame and depth map are first extracted. Then, the master shares and ownership shares are generated based on these features and the watermark information. The extracted features and their relevant ownership shares are stored and injective index relationships between them are established. For a queried 3D video, the similarity-based retrieval phase is first executed in which its features are extracted and compared with the stored features, which is not included in traditional zero-watermark schemes. If any match occurs, the master shares are then generated from the extracted features for copyright identification. Finally, the copyright ownership of the illegal 3D video is identified by stacking the generated master shares with the ownership shares obtained according to the retrieval results.

We also design flexible mechanisms for both similarity-based retrieval and copyright identification to better satisfy the requirements for copyright protection of DIBR 3D videos. When the copyrights of 2D frame and depth map are different, they are retrieved separately, and their copyrights are identified independently. Otherwise, the retrieval and copyright identification results fused following an attention-based fusion method [15] to enhance the DRM performances.

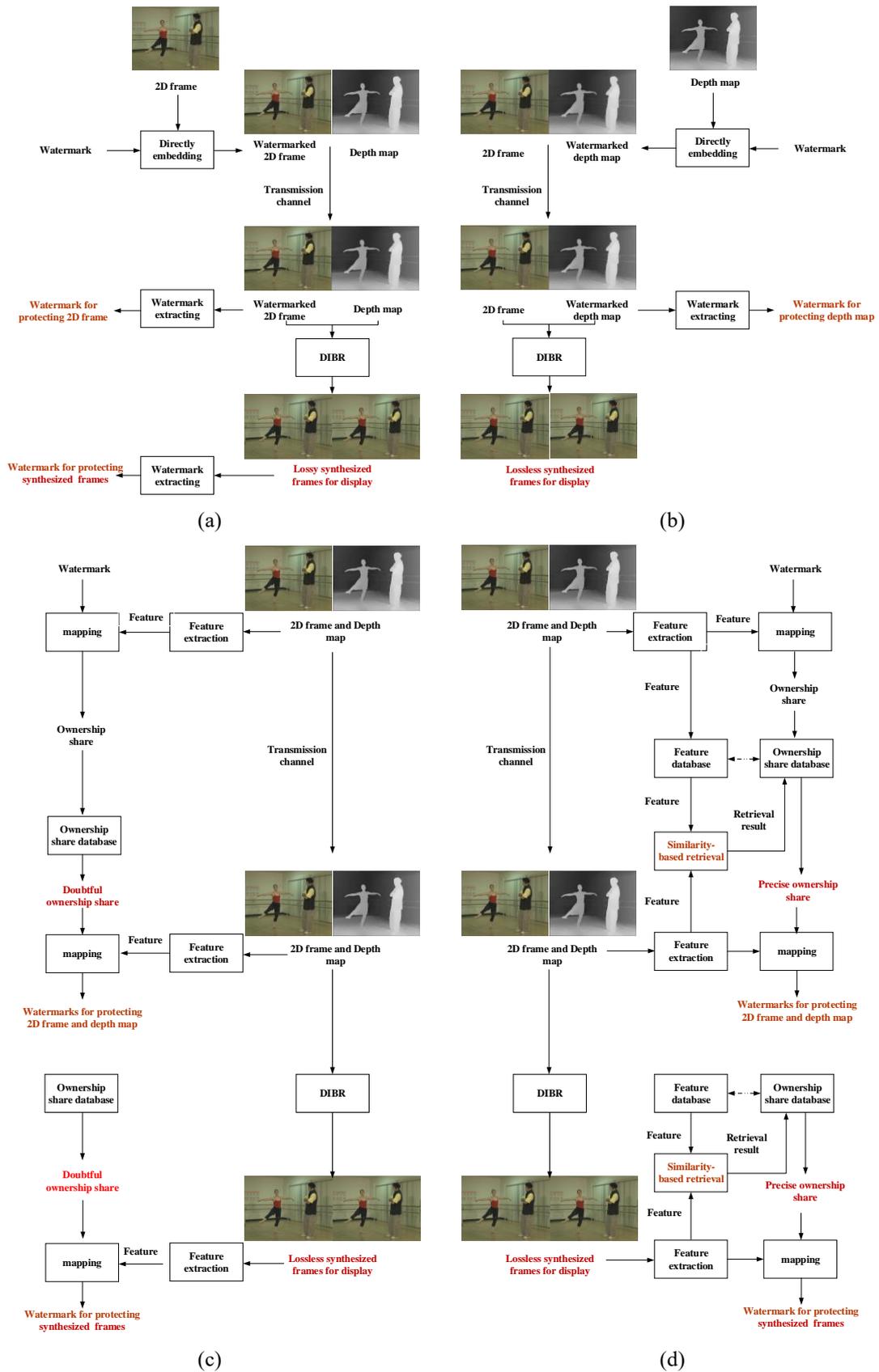

**Fig. 1.** Structure of different schemes: (a) 2D frame-based watermark (b) Depth map-based watermark (c) Zero-watermark (d) Proposed RZW-SR$_{3D}$

Fig. 1 summarizes that how proposed RZW-SR$_{3D}$ differs from existing 2D frame-based watermark, depth map-based watermark and zero-watermark schemes. Moreover, the red-colored part of Fig. 1 emphasizes the differences of these schemes in term of DRM performances and demonstrates the superiority of our proposed RZW-SR$_{3D}$.

Our experimental results demonstrate that the RZW-SR$_{3D}$ have remarkable DRM performances. First, ownership shares relevant to a particular 3D video are obtained precisely and reliably under various video attacks and DIBR distortions even when processing large-scale videos, which outperforms existing zero-watermark scheme. Second, copyrights of both 2D frame and depth map are identified reliably, independently and simultaneously under various video attacks and DIBR distortions without any video distortion, which outperforms existing 2D frame-based and depth map-based watermark schemes.

The rest of the paper is organized as follows. In Section II, the related works are listed. In Section III, the proposed RZW-SR$_{3D}$ is described in detail. The experiment results and analyses are presented in Section IV, and the conclusions are presented in Section V.

## 2. Related work

Existing schemes for copyright protection of DIBR-based 3D video can be mainly classified into three categories, which are 2D frame-based watermark schemes, depth map-based watermark schemes and zero-watermark scheme. Each category is described individually as below.

## 2.1 2D frame-based watermark schemes

2D frame-based watermark schemes [1-3, 6-7, 9, 13, 16, 19, 22, 27, 30, 32, 43] utilize the characteristics of depth maps to achieve remarkable watermark robustness and imperceptibility. Lee et al. [22] proposed a scheme utilizing the areas with high motion on z-axis and the pixels to be hidden by rendering for watermark embedding to ensure the watermark imperceptibility. Fan et al. [13] proposed another scheme, in which depth-perceptual regions of interest (DP-ROI) is constructed by integrating the foreground region, the depth-edge region and the gray-contour region to achieve better watermark imperceptibility. For these schemes, the watermark can be hardly extracted from the synthesized frames, because the pixels are shifted horizontally by DIBR and thus the synchronism of watermark embedding and extracting is destroyed during the DIBR process. To address this issue, lots shift invariant schemes are designed. Kim et al. [16] proposed a scheme by quantizing the coefficients of a dual-tree complex wavelet transform (DT-CWT) to make use of the approximate shift invariance and directional selectivity of DT-CWT. Asikuzzaman et al. [2-3] also proposed DT-CWT based schemes, in which the watermark is embedded in the chrominance channels of 2D frame frames to achieve stronger robustness. Lin et al. [27] proposed a robustness watermark scheme, in which multiple watermarks are embedded and the effects of watermark-embedding order are studied. Niu et al. [7, 43] proposed schemes based on histogram shape, in which suitable groups of pixel histogram are selected for watermark embedding to enhance the watermark robustness against geometric attacks and DIBR distortions. Al-Haj et al. [1] proposed a watermark scheme based on mathematical transforms, in which the discrete wavelet transform (DWT) and the singular value decomposition (SVD) are utilized to provide complementary robustness against watermarking attacks. Chen et al. [9] proposed a scheme in which the watermark is embedded into contourlet subbands of the 2D frame to make use of the directional multiresolution image representation and convenient tree structures of contourlet transform. Lee et al. [30] proposed a scheme in which embedding locations are selected based on depth variation prediction map and watermark are embedded in 2D DWT domain by quantization index modulation (QIM). To further enhance the robustness against geometric attacks, several watermark schemes with geometric resynchronization function [6, 19, 32] are proposed. In these schemes, the resynchronization between watermarked and attacked video frames is based on embedding pre-defined template [19] or utilizing local feature points [6, 32], such as scale invariant feature transform (SIFT) descriptors.

However, there are still some problems of 2D frame-based watermark schemes. First, they introduce some irreversible distortions to the video content which effects the 3D viewing experiences [29]. Second, they only embed watermarks into the 2D frames without considering the situation that generators of 2D frame and depth map of 3D videos may be different. Therefore, they cannot protect the copyright of depth map independently.

## 2.2 Depth map-based watermark schemes

Depth map-based watermark schemes [8, 24-25, 29, 33-34, 38] restrict the modifications of depth maps below a certain threshold to keep the synthesized 3D videos undistorted. One category of depth map-based schemes is based on unseen visible watermark (UVW) technique [10, 14]. Lin et al. [29] and Pei et al. [33-34] proposed several UVW-based watermark schemes. In these schemes, the synthesized 3D videos are undistorted under normal viewing conditions and meaningful watermarks can be recognized under non-normal DIBR conditions. However, UVW-based watermarks are only viewable for subjective copyright identification and can hardly be extracted for objective copyright identification. In addition, the watermark-embedding strength is determined based on prior estimates to restrict the modifications of depth maps in these schemes. Unfortunately, the utilization of prior estimates limits the selection of watermark-embedding methods and thus weakens the watermark robustness. To address these two issues, Liu et al. [25] proposed an advanced unseen extractable watermark (AUEW) scheme. In this scheme, watermarks are embedded by quantizing the sum of AC coefficients in the pseudo 3D-DCT domain and can be directly extracted for objective copyright identification. Moreover, simulations of embedding process rather than prior estimates are utilized to restrict the modifications of depth maps which achieves a better trade-off between watermark robustness and imperceptibility. The other category of depth map-based schemes is depth-no-synthesis-error (D-NOSE)-model based reversible watermark schemes [8, 24, 38]. In these schemes, watermarks are embedded into depth maps based on reversible watermark techniques, such as difference expansion [17], error prediction [39] and histogram modification [20] for lossless recovery of depth maps. Moreover, locations which are suitable for watermarking embedding is selected based on D-NOSE model for better 3D viewing experiences. Although these schemes can ensure that both of depth maps and synthesized 3D videos are lossless, they are vulnerable against attacks and thus cannot provide reliable copyright identification when 3D videos are attacked.

In summary, there are still three problems of depth map-based watermark schemes. First, their robustness against severe signal attacks and geometrical attacks is insufficient [26]. Second, they only embed watermarks into the depth maps but and thus cannot protect the copyright of 2D frame independently. Finally, they cannot extract any copyright information from the synthesized frames for effective copyright protection.

## 2.3 Zero-watermark scheme

Zero-watermark scheme generates mapping ships between video feature and watermark without direct watermark embedding to avoid content-distortion. Liu et al. [26] proposed a robust zero-watermark scheme for DIBR 3D video ($RZW_{3D}$). In this scheme, the master shares are first generated by binarizing the 2D-DCT coefficients of temporally informative representative images (TIRI) [11] of 2D frames and depth maps. Then, the ownership shares are generated based on (2,2) VSS [31] and stored for copyright identification instead of directly embedding watermarks into the video content. The copyrights of a queried 3D video are recovered by stacking the master shares and ownership shares. By utilizing the robust features of both 2D frame and depth map, not only the robustness of zero-watermark scheme against various attacks can be ensured, but also the copyrights of 2D frame and depth map can be identified independently and simultaneously to satisfy the different DRM requirements of DIBR 3D videos.

However, there are still two problems of this zero-watermark scheme. First, $RZW_{3D}$ is based on the premise that the ownership share relevant to a particular video has been already obtained before copyright identification. Otherwise, the copyright of this video cannot be identified correctly and the verification of an illegal copy would fail. Unfortunately, because no efficient indexing mechanism is designed in the zero-watermark scheme, it is difficult to obtain the precise ownership shares when processing large-scale videos. Second, the robustness of $RZW_{3D}$ against geometrical attacks still needs improvements.

## 3. Proposed scheme

The proposed scheme includes three phases, which are a copyright-registration phase, a similarity-based retrieval phase, and a copyright-identification phase, as shown in Fig. 2. Each phase is described individually as below.

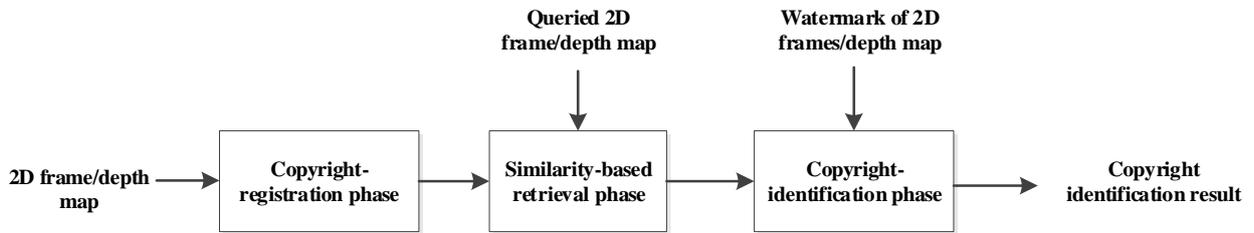

**Fig. 2.** Procedure of our proposed scheme

## 3.1 Copyright-registration phase

In the copyright-registration phase, feature and ownership-share databases of both 2D frame and depth map are constructed from their features and relative watermark information, as shown in Fig. 3.

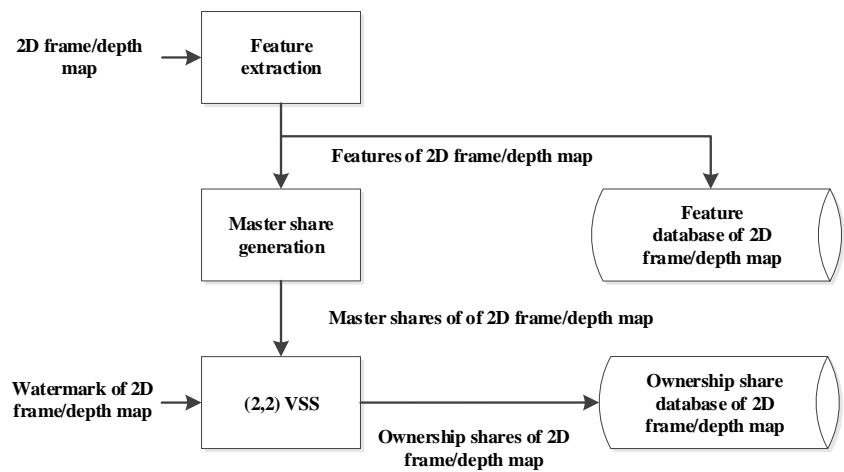

**Fig. 3.** Copyright-registration phase of our proposed scheme

## 3.1.1 Feature extraction

In our study, the same feature-extraction methods of 2D frame and depth map are designed, which reduces computational costs based on reusing benefits. Both the 2D frame and depth map are first smoothed and normalizing in the spatial and temporal domain. Then, temporally informative representative images (TIRI) of these video frames are constructed to exploit the temporal information of the video sequence [11]. After that, deviations between TIRIs and video frames are calculated and normalized. Next, frames consisted of the normalized TIRI-based deviations are partitioned into concentric rings. Finally, centroids of normalized TIRI-based deviations in concentric rings are calculated to generate the final features. The procedure of feature extraction is shown in Fig. 4 and listed below.

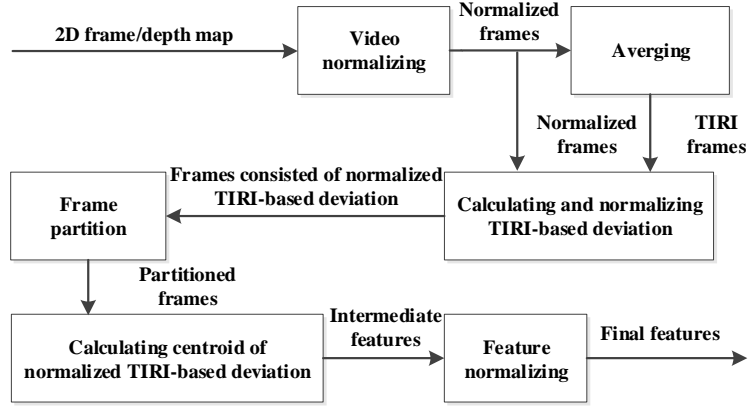

**Fig. 4**. Feature extraction of our proposed scheme

**Step 1.** Normalizing both the 2D frame and depth map, from $w \times h \times l$ to $320 \times 320 \times 100$ based on spatial and temporal resampling and smoothing. Here, $w$ and $h$ are the width and height of the original video frame, respectively. $l$ is the frame number of the 3D video clip. Select the luminance of the normalized 2D frames and depth maps, denoted as $R_{2d}$ and $R_{depth}$ for feature extraction. In this manner, the robustness of extracted features against the resizing and temporal attacks can be enhanced.

**Step 2.** Subsample and average $R_{2d}$ and $R_{depth}$ in temporal domain to construct their TIRIs, denoted as $TIRI_{2d}$ and $TIRI_{depth}$, respectively, in (1).

$$TIRI_c(i, j) = \sum_{m=1}^{M} R_c(i, j, k) \times w_k / \sum_{m=1}^{M} w_k; \qquad (1)$$
$$w_k = a^k, \quad k = 5 \times m.$$

where $c \in (2d, depth)$, $TIRI_{2d}(i, j)$ and $TIRI_{depth}(i, j)$ are the pixels in the $i$ th row and $j$ th column of the 2D frame and depth map TIRIs, respectively. $R_{2d}(i, j, k)$ and $R_{depth}(i, j, k)$ are the pixels in the $i$ th row and $j$ th column of the $k$ th frame of resized 2D frame and depth map; $w_k$ is the weight associated with $R_{2d}(i, j, k)$ and

$R_{depth}(i, j, k)$, $1 \leq i \leq 320, 1 \leq j \leq 320, 1 \leq m \leq 20$, $M=20$ and $a$ is between 0 and 1. When $a$ is close to 1, the generated TIRI is similar to an averaged image and contains more temporal information, which is more robust. Otherwise, the generated TIRI will contain more spatial information and less temporal information, which is more discriminative. In our study, $a$ is selected to be 1. In this manner, the generated TIRI contains more temporal information and thus the feature robustness is stronger against noise addition and temporal attacks.

**Step 3.** Generate TIRI-based deviations $D_{2d}(i, j, k)$ and $D_{depth}(i, j, k)$ by calculating the maximal absolute differences between pixels in normalized frames and their 8 spatial neighborhood pixels in the TIRIs, as shown in (2).

$$D_c(i, j, k) = \max(|TIRI_c(i \pm 1, j \pm 1) - R_c(i, j, k)|); \quad (2)$$

where $c \in (2d, depth)$, $2 \leq i \leq 319, 2 \leq j \leq 319$ and $1 \leq k \leq 100$.

When a video frame suffers attacks, such as rotation or flipping, the 8 spatial neighborhoods of each pixel are still located around the central pixel. Only their relative positions are modified, which does not affect much on the maximal absolute differences calculated in (2). Therefore, the utilization of 8 spatial-neighborhood pixels enhances the robustness against rotation and flipping attacks.

The pixels in TIRIs rather than those in single normalized frames are utilized as the neighborhood pixels to exploit temporal properties of video sequences, which enhances the feature robustness against noise addition and temporal attacks. The pixels in normalized video frames rather than those in TIRIs are utilized as central pixels to exploit more detailed spatial properties of video sequences, which enhances the feature distinguishability.

**Step 4.** Normalize the TIRI-based deviations, denoted as $N_{2d}(i, j, k)$ and $N_{depth}(i, j, k)$, in (3).

$$N_c(i, j, k) = \arctan(D_c(i, j, k) / TIRI_c(i, j)); \quad (3)$$

where $c \in (2d, depth)$. In this manner, the feature robustness against global attacks such as contrast modification, brightness modification and gamma transform is enhanced.

**Step 5.** Partition the frames consisted of normalized TIRI-based deviations into a central circle and $N$-1 concentric rings. The radius of the central circle and the widths of concentric rings are set to $r$.

For each pixel $(i, j, k)$ in the $k$ th frame, its distance $Dist(i, j, k)$ to the center point of this frame $(i_o, j_o, k)$ is first calculated as shown in (4).

$$Dist(i, j, k) = \sqrt{(i - i_o)^2 + (j - j_o)^2} \quad (4)$$

Then, the partition $n$ of pixel $(x, y, k)$ is calculated based on $Dist(i, j, k)$, as shown in (5).

$$n = \lfloor Dist(i, j, k) / r \rfloor \quad (5)$$

In our study, the size of frames consisted of normalized TIRI-based deviations is 320×320, $N$ is equal to 16, $r$ is equal to 10. When the videos are rotated or flipped, the pixels partitioned in this manner still belong to their original associated circle or ring partitions. Therefore, feature robustness against rotation and flipping attacks is ensured. In addition, the regions outside the largest ring are not utilized in our study due to the following two reasons. On one hand, the main visual of a video frame are usually concentrated in its central region, and the importance of a pixel increases as its distance to the frame center decreases in common cases. Therefore, features generated by discarding the regions outside the largest ring do not lose much important visual information. On the other hand, because these regions are the most common places for logo insertion and edge cropping attacks, the feature robustness against these attacks can be enhanced by discarding the regions outside the largest ring.

**Step 6.** Take pixel values in TIRIs as the weights of normalized TIRI-based deviations. Calculate centroids of normalized TIRI-based deviations in each partition using (6) to generate the intermediate features of 2D frame and depth map, denoted as $f_{2d}$ and $f_{depth}$, using (7).

$$v_c(n,k) = \frac{\sum_{(n-1)r \leq Dist(i,j,k) \leq nr} TIRI_c(i,j) N_c(i,j,k)}{\sum_{(n-1)r \leq Dist(i,j,k) \leq nr} TIRI_c(i,j)}; \quad (6)$$

$$f_c = [v_c(1,1) \ ... \ v_c(N,1) \ ... \ v_c(1,K) \ ... \ v_c(N,K)]; \quad (7)$$

where $c \in (2d, depth)$, $K$ is the number of frames consisted of normalized TIRI-based deviations. In our study, $K$ is equal to 100, and the dimensionality of $f_{2d}$ and $f_{depth}$ is $K \times N$ equal to 1600.

**Step 7.** Normalize the intermediate features by their mean and standard deviation as shown in (8) to generate the final features, $fn_{2d}$ and $fn_{depth}$. Store $fn_{2d}$ and $fn_{depth}$ in the 2D frame and depth map feature databases, respectively, for similarity-based retrieval and copyright identification. Here, $c \in (2d, depth)$.

$$\mu_c = \frac{1}{NK} \sum_{i=1}^{NK} f_c(i);$$
$$\sigma_c = \sqrt{\frac{1}{NK-1} \sum_{i=1}^{NK} (f_c(i) - \mu_c)^2}; \quad (8)$$
$$fn_c(i) = \frac{f_c(i) - \mu_c}{\sigma_c};$$

## 3.1.2 Generation of master share and ownership share

$(m, n)$ VSS (where $m \leq n$) was proposed by Naor *et al*. [31]. In VSS, one binary image is split into $n$ shares, and can be recovered from $l$ shares when $l \geq m$. Otherwise, the image cannot be recovered. The (2, 2) VSS is a typical VSS, in which each pixel of a binary image is substituted by a pair of shares consisting of four sub-pixels. A white pixel is split into two identical shares, whereas a black pixel is split into two complementary shares, as shown in Table 1.

(2-2) VSS is a low-cost injective function and utilized in our RZW-SR$_{3D}$ to generate the master shares and ownership shares. In this manner, mapping relationships between features and copyright information is injective. Therefore, robustness and distinguishability of features can be maintained during the procedure of generating master shares and ownership shares to achieve remarkable DRM performance. The detailed procedure of generating master shares and ownership shares is listed below.

**Table 1.** Typical VSS: (2, 2) VSS

| Pixel value | 1 (white pixel) | 0 (black pixel) |
|---|---|---|
| Master share | | |
| Ownership share | | |
| Stack result | | |

**Step 1.** Binarize the extracted features of 2D frame and depth map by their median values to construct intermediate vectors, as shown in (9).

$$I_c(i) = 1, \ if \ fn_c(i) > t_c;$$
$$I_c(i) = 0, \ if \ fn_c(i) \le t_c \tag{9}$$

where $c \in (2d, depth)$, $1 \le i \le K \times N$, $t_{2d}$ is the median value of $f_{2d}$, and $t_{depth}$ is the median value of $f_{depth}$.

**Step 2.** Rearrange the intermediate vectors $I_{2d}$ and $I_{depth}$ to construct intermediate matrices $V_{2d}$ and $V_{depth.}$, which are of size 40×40. Generate the master shares of 2D frame and depth map, which are denoted as $M_{2d}$ and $M_{depth}$, according to the (2, 2) VSS as shown in (10).

$$m_c(i, j) = \begin{cases} \begin{bmatrix} 1,0 \\ 0,1 \end{bmatrix}, V_c(i, j) = 1; \\ \begin{bmatrix} 0,1 \\ 1,0 \end{bmatrix}, V_c(i, j) = 0; \end{cases} \tag{10}$$

where $c \in (2d, depth)$, $m_{2d}(i, j)$ and $m_{depth}(i, j)$ are non-overlapping 2 × 2 blocks of $M_{2d}$ and $M_{depth}$; $1 \le i \le 40$ and $1 \le j \le 40$.

**Step 3.** Generate ownership shares of 2D frame and depth map, which are denoted as $O_{2d}$ and $O_{depth}$, according to the (2, 2) VSS as shown in (11).

$$o_c(i,j) = \begin{cases} \begin{bmatrix} 1,0 \\ 0,1 \end{bmatrix}, \text{if } m_c(i,j) = \begin{bmatrix} 1,0 \\ 0,1 \end{bmatrix}, W_c(i,j) = 1; \\ \begin{bmatrix} 0,1 \\ 1,0 \end{bmatrix}, \text{if } m_c(i,j) = \begin{bmatrix} 1,0 \\ 0,1 \end{bmatrix}, W_c(i,j) = 0; \\ \begin{bmatrix} 0,1 \\ 1,0 \end{bmatrix}, \text{if } m_c(i,j) = \begin{bmatrix} 0,1 \\ 1,0 \end{bmatrix}, W_c(i,j) = 1; \\ \begin{bmatrix} 1,0 \\ 0,1 \end{bmatrix}, \text{if } m_c(i,j) = \begin{bmatrix} 0,1 \\ 1,0 \end{bmatrix}, W_c(i,j) = 0; \end{cases} \quad (11)$$

where $c \in (2d, depth)$, $o_{2d}(i,j)$ and $o_{depth}(i,j)$ are non-overlapping $2 \times 2$ blocks of $O_{2d}$ and $O_{depth}$; $W_{2d}(i,j)$ and $W_{depth}(i,j)$ are the watermark bits of 2D frame and depth map; $1 \le i \le 40$ and $1 \le j \le 40$.

**Step 4.** Store the generated $O_{2d}$ and $O_{depth}$ in ownership databases and establish injective index relationships between the extracted features and their relevant ownership shares for copyright-identification phase.

## 3.2 Similarity-based retrieval phase

When a 3D video is queried, a similarity-based retrieval phase is executed to find the stored features that match the features extracted from the queried video, as shown in Fig. 5.

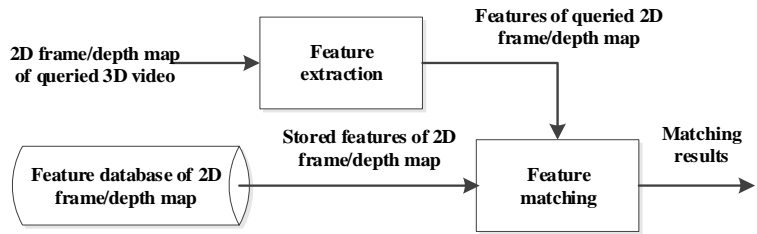

**Fig. 5.** Similarity-based retrieval phase of our proposed scheme

Features of queried 2D frame and depth map, which are denoted as $fn'_{2d}$ and $fn'_{depth}$, are first extracted following the same procedure in the copyright-registration phase. Then, normalized distances between queried features and original features in databases are measured as shown in (12).

$$d_c = \frac{1}{NK} \sum_{i=1}^{NK} (fn_c(i) - fn'_c(i))^2; \quad (12)$$

where $c \in (2d, depth)$, $N$ is the number of partitions and equal to 16, and $K$ is the number of frames and equal to 100.

To satisfy different DRM requirements of DIBR 3D videos, a flexible matching mechanism is designed in our study. In the case that the copyrights of 2D frame and depth map are different, the features of 2D frame and depth map are matched separately and independently. Given the thresholds $T_{2d}$ and $T_{depth}$, if any $d_{2d}$ or $d_{depth}$ is smaller than $T_{2d}$ or $T_{depth}$, respectively, the 2D frames or depth maps of two DIBR 3D videos are considered as near-duplicates and the queried DIBR 3D video is considered as illegal. Otherwise, normalized distances of 2D frame features and depth map features are fused. It will be better if the function $F_D$ satisfies the heterogeneity and monotonicity [15] for distance fusion, which are defined in (13)-(14), respectively.

$$F_D(d_1, d_2) > F_D(d_1 - \varepsilon, d_2 + \varepsilon) \tag{13}$$

where $0 < \varepsilon \leq d_1 \leq d_2$, $d_1$, $d_2$ are distance values to be fused.

$$\begin{aligned} F_D(d_1, d_2) &< F_D(d_1 + \varepsilon, d_2); \\ F_D(d_1, d_2) &< F_D(d_1, d_2 + \varepsilon). \end{aligned} \tag{14}$$

where $0 < \varepsilon \leq d_1 \leq d_2$; $d_1$, $d_2$ are distance values to be fused.

To strictly satisfy the heterogeneity and monotonicity simultaneously, an attention-based fusion function for distances is designed inspired by the method [15] as below.

$$\begin{aligned} x_1 &= \frac{1}{d_{2d}} + \frac{1}{d_{depth}}; \\ x_2 &= \left| \frac{1}{d_{2d}} - \frac{1}{d_{depth}} \right| \\ F_D(d_{2d}, d_{depth}) &= \frac{1}{\frac{1}{2}(x_1 + \frac{1}{1+\gamma} x_2)}. \end{aligned} \tag{15}$$

where $\gamma$ is a constant and is set to 0.1 empirically in our study.

Given a threshold $T_{fusion}$, the two DIBR 3D videos are considered near-duplicates and the queried DIBR 3D video is considered as illegal if any distance value fused in (15) is smaller than $T_{fusion}$.

Because the 2D frames have more texture information than the depth maps whereas the distribution of pixel values in the depth maps is regional, robustnesses of 2D frame and depth map features are complementary to a certain extent. Therefore, the distance fused in (15) further enhances the reliability of similarity-based retrieval against different attacks.

If the queried DIBR 3D video is considered as illegal, a copyright-identification phase is implemented. Otherwise, the procedure of RZW-SR$_{3D}$ is finished.

## 3.3 Copyright-identification phase

In this phase, ownership shares relevant to the matched features are utilized along with the master shares generated from illegal DIBR 3D videos to recover the copyright information as shown in Fig. 6. The detailed procedure is listed below.

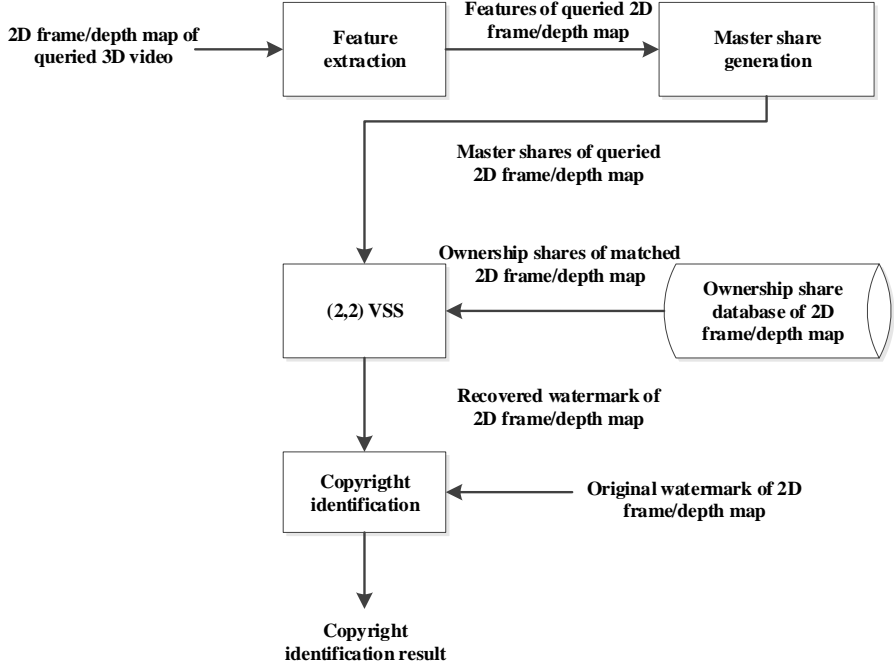

**Fig. 6.** Copyright-identification phase of our proposed scheme

## 3.3.1 Recovery of copyright information

**Step 1.** Obtain the ownership shares relevant to the matched features from ownership databases.

**Step 2.** Generate the master shares of illegal DIBR 3D videos, $M'_{2d}$ and $M'_{depth}$, from $f'_{2d}$ and $f'_{depth}$, respectively, following the same procedure in the copyright-registration phase.

**Step 3.** Stack the generated master shares with the obtained ownership shares to construct intermediate copyright-identification matrices $S_{2d}$ and $S_{depth}$, as shown in Table 1.

**Step 4.** Recover the watermark information of the 2D frame and depth map $W'_{2d}$ and $W'_{depth}$, from $S_{2d}$ and $S_{depth}$, respectively, in (16).

$$W'_c(i,j) = \begin{cases} 1, & \sum s_c^{i,j}(x,y) \geq 2; \\ 0, & \sum s_c^{i,j}(x,y) < 2; \end{cases} \quad (16)$$

where $c \in (2d, depth)$; $s_{2d}^{i,j}(x,y)$ and $s_{depth}^{i,j}(x,y)$ are non-overlapping 2×2 blocks of $S_{2d}$ and $S_{depth}$; $1 \leq i \leq 40$, $1 \leq j \leq 40$, $1 \leq x \leq 2$, and $1 \leq y \leq 2$.

## 3.3.2 Identification of copyright ownership

To satisfy the different DRM requirements of DIBR 3D videos, a flexible copyright-identification mechanism is designed in our study.

In the case that the copyrights of 2D frame and depth map are different, they are identified separately and independently by calculating their own bit error rates (*BER*) between the original watermark and the recovered watermark as shown in (17).

$$BER_c = \frac{\sum W_c(i,j) \oplus W_c^{'}(i,j)}{i \times j};  \quad (17)$$

where $c \in (2d, depth)$, $1 \leq i \leq 40$ and $1 \leq j \leq 40$.

Otherwise, the *BER* values of 2D frame and depth map are fused. It will be better if the function $F_{BER}$ satisfies the heterogeneity and monotonicity for *BER* fusion, which are similar with distance fusion. Therefore, a similar attention-based fusion function for *BER* is also designed as below,

$$x_1 = \frac{1}{BER_{2d}} + \frac{1}{BER_{depth}};$$
$$x_2 = |\frac{1}{BER_{2d}} - \frac{1}{BER_{depth}}| \quad (18)$$

$$F_{BER}(BER_{2d}, BER_{depth}) = \frac{1}{\frac{1}{2}(x_1 + \frac{1}{1+\gamma}x_2)}. \quad (19)$$

where $\gamma$ is a constant set to 0.1 empirically in our study.

The attention-based fusion further enhances the reliability of copyright identification because the complementary robustnesses of 2D frame and depth map features as analyzed in the similarity-based retrieval phase.

## 4 Experimental Result and Discussion

## 4.1 Experimental Setup

In this study, performances of similarity-based retrieval in RZW-SR$_{3D}$ are first evaluated in section 4.2. Then, performances of copyright identification in RZW-SR$_{3D}$ are evaluated in section 4.3. Finally, qualitative comparisons between RZW-SR$_{3D}$ and other schemes for DIBR 3D videos are represented in section 4.4. Our testing database contains 200 different DIBR 3D video clips. Some of them are from the open datasets provided by the MEPG 3DAV group [12] and Interactive Visual Media Group of Microsoft Research [45]. Others consist of 2D frames selected from existing movies, with their depth maps manually calibrated by authors based on the method in [37]. The video clips in [12] consist 125 or 250 frames at 25 fps and those in [45] consist 100 frames at 15 fps. Their frame sizes are 720 × 576 and 1024 × 768 respectively. For our manually generated DIBR video clips, they consist 200 frames at 25 fps and their frame size are 1920 ×

1080. A watermark image of size 40 × 40 is utilized for copyright identification. 4 sequences of synthesized frames are generated by set the baseline distances as 5% and 7% of the video frame width. These baseline distances have been chosen carefully because baseline distances larger than 7% may induce discomfort for 3D viewing [1, 36].

All original 2D frames, depth maps and synthesized frame suffer 14 types, totally 26 different attacks. As a result, a total of 32000 sequences of queried videos, including 5200 sequences of attacked 2D frames, 5200 sequences of attacked depth maps, 800 sequences of original synthesized frames and 20800 sequences of attacked synthesized frames, are generated to evaluate performances of the retrieval and copyright identification functions of RZW-SR$_{3D}$. The types and parameters of the attacks are shown in Table 2. The 1$^{st}$ to 8$^{th}$ are global attacks. The 9$^{th}$ to 12$^{th}$ are geometrical attacks. The last two are temporal attacks.

**TABLE 2.** Types of attacks

| Types | Parameters |
|---|---|
| Gaussian blurring (GB) | window=9×9, variance=1; window=15×15, variance=1 |
| Average filtering (AF) | window=9×9; 15×15 |
| Median filtering (MF) | window=9×9; 15×15 |
| Changes in contrast (CC) | -30%; +30% |
| Changes in brightness (CB) | -30%; +30% |
| Gamma transform (GT) | γ =0.6; 1.4 |
| Gaussian noise addition (GN) | Mean=0, variance=0.005; Mean=0, variance=0.01 |
| Logo insertion in the upper-left (LI) | Logo size =32 × 32; 64×64 |
| Resizing (RS) | 1/2 of original size; 1/5 of original size |
| Crop from image edges (CR) | 5%; 10% |
| Rotation (RT) | 45°; 90° |
| Flip (FL) | Vertical, horizontal |
| Frame replacing randomly (FR) | Replacing rate=5% |
| Frame dropping randomly (FD) | Dropping rate=5% |

## 4.2 Evaluation of similarity-based retrieval performance

Performances of similarity-based retrieval is evaluated by calculating the false-negative rate $P_{fn}$s with a fixed false-positive rate $P_{fp}$s under all these attacks. $P_{fp}$ is a probability of classifying different videos as the same and indicates feature distinguishability, whereas $P_{fn}$ is a probability of classifying original and its attack videos to be different and indicates feature robustness. $P_{fp}$ and $P_{fn}$ are defined by (20)-(21).

$$P_{fp} = \frac{N_{fp}}{N_{dis}} \tag{20}$$

$$P_{fn} = \frac{N_{fn}}{N_s} \qquad (21)$$

where $N_{fp}$ is the number of different video-clip pairs whose Hamming distance is smaller than a threshold, $N_{dis}$ is the true number of different video clips, $N_{fn}$ is the number of pairs of original and its attack videos whose Hamming distance is larger than the threshold, and $N_s$ is the true number of attacked videos.

First, performances of similarity-based retrieval in our proposed RZW-SR$_{3D}$ when utilizing original 2D frames, utilizing depth maps and fusing them together are tested on all the DIBR 3D videos in our databases. The results are listed in Table 3.

**TABLE 3.** $P_{fn}$s with $P_{fp}$s =0.01 under different attacks when utilizing original 2D frames, depth maps and fusing them

|  | RZW-SR$_{3D}$ | | |
| --- | --- | --- | --- |
|  | 2D frame | Depth map | Two together |
| GF 9×9 | 0.0000 | 0.0200 | 0.0000 |
| GF 15×15 | 0.0000 | 0.0200 | 0.0000 |
| AF 9×9 | 0.0100 | 0.0350 | 0.0000 |
| AF 15×15 | 0.0100 | 0.0450 | 0.0000 |
| MF 9×9 | 0.0100 | 0.0000 | 0.0000 |
| MF15×15 | 0.0100 | 0.0000 | 0.0000 |
| CC -30 | 0.0000 | 0.0000 | 0.0000 |
| CC +30 | 0.0000 | 0.0150 | 0.0000 |
| CB -30 | 0.0100 | 0.0000 | 0.0000 |
| CB +30 | 0.0000 | 0.0400 | 0.0000 |
| GT γ =0.6 | 0.0000 | 0.0000 | 0.0000 |
| GT γ =1.4 | 0.0000 | 0.0000 | 0.0000 |
| GN 0.005 | 0.0050 | 0.0650 | 0.0000 |
| GN 0.01 | 0.0050 | 0.0900 | 0.0050 |
| LI 32×32 | 0.0000 | 0.0000 | 0.0000 |
| LI 64×64 | 0.0000 | 0.0400 | 0.0000 |
| RS 1/2 | 0.0000 | 0.0000 | 0.0000 |
| RS 1/5 | 0.0100 | 0.0050 | 0.0000 |
| CR 5% | 0.0000 | 0.1300 | 0.0100 |
| CR 10% | 0.0250 | 0.1450 | 0.0150 |
| RT 45° | 0.0050 | 0.1050 | 0.0000 |
| RT 90° | 0.0350 | 0.1450 | 0.0250 |
| FL Vertical | 0.0000 | 0.0000 | 0.0000 |
| FL Horizontal | 0.0000 | 0.0000 | 0.0000 |
| FR 5% | 0.0100 | 0.1450 | 0.0050 |
| FD 5% | 0.0350 | 0.0250 | 0.0150 |
| Average value | 0.0067 | 0.0412 | 0.0029 |

**TABLE 4.** $P_{fn}$s with $P_{fp}$s =0.01 under different attacks when utilizing depth maps, synthesized frames, and fusing them

| | Depth map | Baseline distance=5% frame width | | | | Baseline distance=7% frame width | | | |
|---|---|---|---|---|---|---|---|---|---|
| | | Synthesized frame (left) | Two together | Synthesized frame (right) | Two together | Synthesized frame (left) | Two together | Synthesized frame (right) | Two together |
| Without attack | 0.0000 | 0.0000 | 0.0000 | 0.0000 | 0.0000 | 0.0100 | 0.0000 | 0.0000 | 0.0000 |
| GB 9×9 | 0.0200 | 0.0000 | 0.0000 | 0.0000 | 0.0000 | 0.0100 | 0.0000 | 0.0000 | 0.0000 |
| GB 15×15 | 0.0200 | 0.0000 | 0.0000 | 0.0000 | 0.0000 | 0.0100 | 0.0000 | 0.0000 | 0.0000 |
| AF 9×9 | 0.0350 | 0.1500 | 0.0000 | 0.0100 | 0.0000 | 0.0200 | 0.0000 | 0.0150 | 0.0000 |
| AF 15×15 | 0.0450 | 0.0200 | 0.0000 | 0.0100 | 0.0000 | 0.0250 | 0.0000 | 0.0150 | 0.0000 |
| MF 9×9 | 0.0000 | 0.0100 | 0.0000 | 0.0100 | 0.0000 | 0.0200 | 0.0000 | 0.0150 | 0.0000 |
| MF15×15 | 0.0000 | 0.0200 | 0.0000 | 0.0100 | 0.0000 | 0.0200 | 0.0000 | 0.0150 | 0.0000 |
| CC -30 | 0.0000 | 0.0000 | 0.0000 | 0.0000 | 0.0000 | 0.0100 | 0.0000 | 0.0000 | 0.0000 |
| CC +30 | 0.0150 | 0.0050 | 0.0000 | 0.0000 | 0.0000 | 0.0150 | 0.0000 | 0.0000 | 0.0000 |
| CB -30 | 0.0000 | 0.0100 | 0.0000 | 0.0050 | 0.0000 | 0.0200 | 0.0000 | 0.0050 | 0.0000 |
| CB +30 | 0.0400 | 0.0000 | 0.0000 | 0.0000 | 0.0000 | 0.0100 | 0.0000 | 0.0000 | 0.0000 |
| GT γ =0.6 | 0.0000 | 0.0000 | 0.0000 | 0.0000 | 0.0000 | 0.0100 | 0.0000 | 0.0000 | 0.0000 |
| GT γ =1.4 | 0.0000 | 0.0000 | 0.0000 | 0.0000 | 0.0000 | 0.0100 | 0.0000 | 0.0000 | 0.0000 |
| GN 0.005 | 0.0650 | 0.0050 | 0.0000 | 0.0000 | 0.0000 | 0.0150 | 0.0000 | 0.0000 | 0.0000 |
| GN 0.01 | 0.0900 | 0.0100 | 0.0050 | 0.0050 | 0.0050 | 0.0200 | 0.0100 | 0.0050 | 0.0100 |
| LI 32×32 | 0.0000 | 0.0000 | 0.0000 | 0.0000 | 0.0000 | 0.0100 | 0.0000 | 0.0000 | 0.0000 |
| LI 64×64 | 0.0400 | 0.0000 | 0.0000 | 0.0000 | 0.0000 | 0.0050 | 0.0000 | 0.0000 | 0.0000 |
| RS 1/2 | 0.0000 | 0.0000 | 0.0000 | 0.0000 | 0.0000 | 0.0100 | 0.0000 | 0.0000 | 0.0000 |
| RS 1/5 | 0.0050 | 0.0100 | 0.0000 | 0.0100 | 0.0000 | 0.0200 | 0.0000 | 0.0150 | 0.0000 |
| CR 5% | 0.1300 | 0.0050 | 0.0100 | 0.0150 | 0.0150 | 0.0250 | 0.0100 | 0.0150 | 0.0150 |
| CR 10% | 0.1450 | 0.0200 | 0.0150 | 0.0200 | 0.0150 | 0.0400 | 0.0150 | 0.0300 | 0.0150 |
| RT 45 | 0.1050 | 0.0150 | 0.0100 | 0.0100 | 0.0150 | 0.0300 | 0.0150 | 0.0250 | 0.0200 |
| RT 90 | 0.1450 | 0.0500 | 0.0300 | 0.0600 | 0.0400 | 0.0550 | 0.0350 | 0.0750 | 0.0500 |
| FL Vertical | 0.0000 | 0.0000 | 0.0000 | 0.0000 | 0.0000 | 0.0000 | 0.0000 | 0.0050 | 0.0000 |
| FL Horizontal | 0.0000 | 0.0000 | 0.0000 | 0.0000 | 0.0000 | 0.0100 | 0.0000 | 0.0000 | 0.0000 |
| FR 5% | 0.1450 | 0.0200 | 0.0100 | 0.0100 | 0.0100 | 0.0300 | 0.0250 | 0.0150 | 0.1500 |
| FD 5% | 0.0250 | 0.0350 | 0.0150 | 0.0400 | 0.0150 | 0.0500 | 0.0150 | 0.0600 | 0.1500 |
| Average value | 0.0412 | 0.0092 | 0.0035 | 0.0079 | 0.0042 | 0.0189 | 0.0046 | 0.0114 | 0.0052 |

Table 3 shows that all the $P_{fn}$s of our proposed RZW-SR$_{3D}$ when utilizing original 2D frames, utilizing depth maps and fusing them together are very small, with their maximal values as 0.0350, 0.1450, 0.0250, and average values as 0.0067, 0.0412, 0.0029. These results demonstrate that the similarity-based retrieval of our proposed RZW-SR$_{3D}$ is effective. In addition, $P_{fn}$s obtained when fusing the two components together are even smaller than those obtained when only utilizing original 2D frames or depth maps, which demonstrate that the attention-based fusion enhances the performances of similarity-based retrieval. The reason for this phenomenon is that the feature robustnesses of 2D frame and depth map are complementary due to 2D frames have more texture information whereas depth maps, of which the pixel values are distributed regionally, are much smoother.

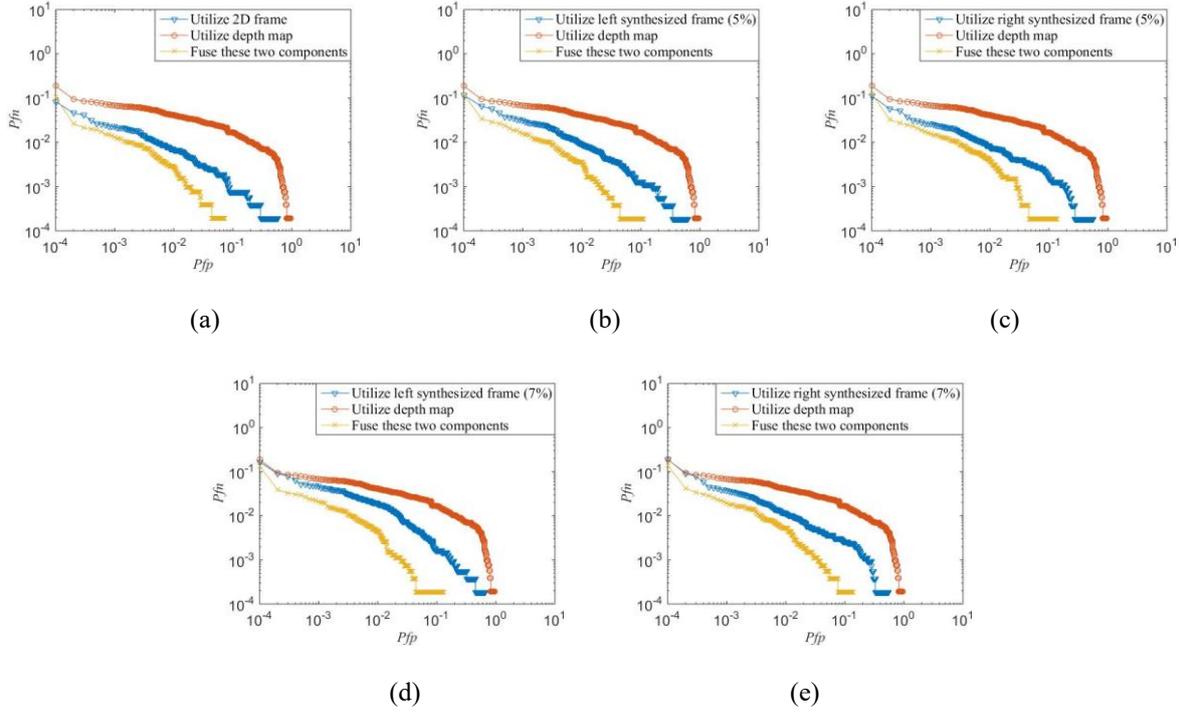

**Fig. 7.** DET graph of the proposed RZW-SR$_{3D}$, (a) utilize original 2D frames, utilize depth maps and fuse them (b) utilize left synthesized frames (baseline=5%), utilize depth maps and fuse them (c) utilize right synthesized frames (baseline=5%), utilize depth maps and fuse them (d) utilize left synthesized frames (baseline=7%), utilize depth maps and fuse them (e) utilize right synthesized frames (baseline=7%), utilize depth maps and fuse them.

Then, performances of similarity-based retrieval in our proposed RZW-SR$_{3D}$ when utilizing the synthesized frames and fusing them with their depth maps are also tested. The results are listed in Table 4.

Table 4 shows that all the $P_{fn}$s of our proposed RZW-SR$_{3D}$ when utilizing synthesized frames are still very small, with their maximal values as 0.0500, 0.0600, 0.0550, 0.0750 respectively, and their average values as 0.0092, 0.0079, 0.0189, 0.0114. Especially, for the synthesized frames without attacks, their $P_{fn}$s are 0.0000, 0.0000, 0.0100 and 0.0000. These results demonstrate that the similarity-based retrieval of our proposed RZW-SR$_{3D}$ when utilizing synthesized frames is still effective although it is slightly worse than utilizing original 2D frames. The reason for this phenomenon is the similarity among the features extracted from the original 2D frames and synthesized frames. In addition, $P_{fn}$s obtained when fusing the synthesized frame and depth maps are still smaller than those when only utilizing synthesized frames or depth maps, with their maximal values as 0.0300, 0.0400, 0.0350, 0.0500 and average values as 0.0035, 0.0042, 0.0046, and 0.0052, which is consistent with the results listed in Table 3.

Finally, we plot a detection error tradeoff (DET) graphs of RZW-SR$_{3D}$, as shown in Fig.7, to fully demonstrate performances of similarity-based retrieval. Fig.7 shows that mean $P_{fn}$s are still not high, and less than 0.2 even when the $P_{fp}$s are set to $10^{-4}$. In addition, all the DET graphs fusing the 2D frame and depth map are below the graphs when only utilizing 2D frame or depth map respectively. These results demonstrate that both the accuracy and the reliability of the similarity-based retrieval in RZW-SR$_{3D}$ are sufficient. In addition, the performance of the similarity-based retrieval is even enhanced by designing the attention-based fusion when the copyrights of 2D frame and depth map are the same.

Because injective index relationships between the features and ownership shares are established, the ownership share relevant to a particular DIBR 3D video can be also obtained precisely and reliably according to similarity-based retrieval results for the latter copyright identification phase.

## 4.3 Evaluation of copyright-identification performance

The performance of copyright identification is mainly in terms of accuracy and reliability. Because only the DIBR 3D video which exists matching features would enter the copyright identification phase, the accuracy of copyright identification has been ensured in the similarity-based retrieval phase by setting $P_{fp}$s to 0.01. Therefore, we mainly evaluate the reliability of copyright identification by using our proposed RZW-SR$_{3D}$ in this section. Besides, we mainly compare our proposed scheme with existing distortion-free schemes for DIBR 3D videos because it is unfair to compare lossless watermark schemes with lossy ones.

A subjective evaluation is first performed. The recovered watermark images in our proposed RZW-SR$_{3D}$ for a typical DIBR 3D video clip 'Ballet' from the open dataset [45] are compared with those detected by UVW [34] and AUEW [25]. The results are shown in Table 5. RZW$_{3D}$ [26] is not included in the subjective evaluation because the copyright information in RZW$_{3D}$ is features of watermark images rather than watermark images themselves, which can hardly be recognized subjectively.

As shown in Table 5, the watermarks recovered from the original 2D frames, depth maps, and synthesized frames by using RZW-SR$_{3D}$ can be recognized successfully under all the attacks for copyright identifications. In addition, they are much clearer than those detected by the other two schemes. The watermarks in UVW are only viewable but can be hardly extracted. Moreover, the watermarks in the UVW can be hardly recognized under filtering, noise addition, severe resizing and rotation attacks. Meanwhile, the watermarks extracted in AUEW can be hardly recognized under severe filtering, heavy noise addition, rotation and flipping attacks. These results demonstrate that our proposed RZW-SR$_{3D}$ can identify the copyright reliably under various attacks and outperforms the other two watermark schemes in terms of the watermark robustness.

**TABLE 5.** Subjective evaluation of copyright-identification performance

| | UVW [34] | AUEW [25] | RZW-SR$_{3D}$ 2D frame | RZW-SR$_{3D}$ Depth map | | UVW [34] | AUEW [25] | RZW-SR$_{3D}$ 2D frame | RZW-SR$_{3D}$ Depth map |
|---|---|---|---|---|---|---|---|---|---|
| Embedded watermark | | | | | Without attack | | | | |
| GB 9×9 | | | | | GB 15×15 | | | | |
| AF 9×9 | | | | | AF 15×15 | | | | |
| MF 9×9 | | | | | MF 15×15 | | | | |
| CC -30 | | | | | CC +30 | | | | |
| CB -30 | | | | | CB +30 | | | | |
| GT γ =0.6 | | | | | GT γ =1.4 | | | | |
| GN 0.005 | | | | | GN 0.01 | | | | |
| LI 32×32 | | | | | LI 64×64 | | | | |
| RS 1/2 | | | | | RS 1/5 | | | | |
| CR 5% | | | | | CR 10% | | | | |
| RT 45° | | | | | RT 90° | | | | |
| FL Vertical | | | | | FL Horizontal | | | | |
| FR 5% | | | | | FD 5% | | | | |

| | RZW-SR$_{3D}$ (baseline distance=5%) | | RZW-SR$_{3D}$ (baseline distance=7%) | | | RZW-SR$_{3D}$ (baseline distance=5%) | | RZW-SR$_{3D}$ (baseline distance=7%) | |
|---|---|---|---|---|---|---|---|---|---|
| | Synthesized frame (left) | Synthesized frame (right) | Synthesized frame (left) | Synthesized frame (right) | | Synthesized frame (left) | Synthesized frame (right) | Synthesized frame (left) | Synthesized frame (right) |
| Embedded watermark | | | | | Without attack | | | | |
| GB 9×9 | | | | | GB 15×15 | | | | |
| AF 9×9 | | | | | AF 15×15 | | | | |
| MF 9×9 | | | | | MF 15×15 | | | | |

**TABLE 5.** Subjective evaluation of copyright-identification performance (continued)

| | RZW-SR$_{3D}$ (baseline distance=5%) | | RZW-SR$_{3D}$ (baseline distance=7%) | | | RZW-SR$_{3D}$ (baseline distance=5%) | | RZW-SR$_{3D}$ (baseline distance=7%) | |
|---|---|---|---|---|---|---|---|---|---|
| | Synthesized frame (left) | Synthesized frame (right) | Synthesized frame (left) | Synthesized frame (right) | | Synthesized frame (left) | Synthesized frame (right) | Synthesized frame (left) | Synthesized frame (right) |
| CC -30 | 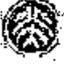 | 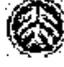 | 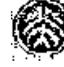 | 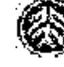 | CC +30 | 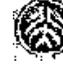 | 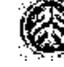 |  |  |
| CB -30 | 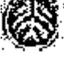 | 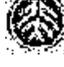 | 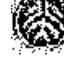 | 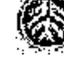 | CB +30 | 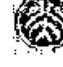 | 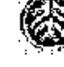 | 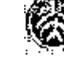 | 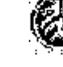 |
| GT γ =0.6 | 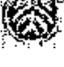 | 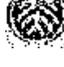 | 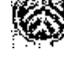 | 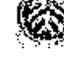 | GT γ =1.4 | 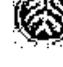 | 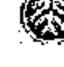 | 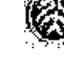 | 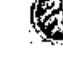 |
| GN 0.005 | 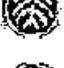 | 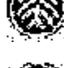 | 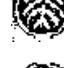 | 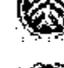 | GN 0.01 | 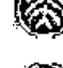 | 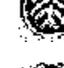 | 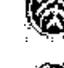 | 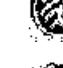 |
| LI 32×32 | 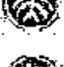 | 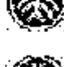 | 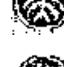 | 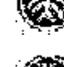 | LI 64×64 | 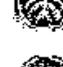 | 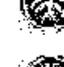 | 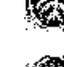 | 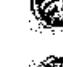 |
| RS 1/2 | 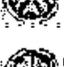 | 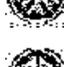 | 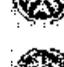 | 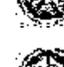 | RS 1/5 | 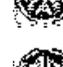 | 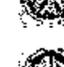 | 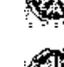 | 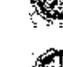 |
| CR 5% | 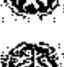 | 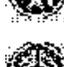 | 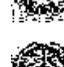 | 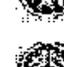 | CR 10% | 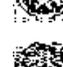 | 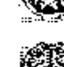 | 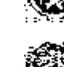 | 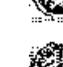 |
| RT 45° | 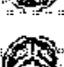 | 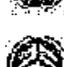 | 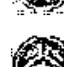 | 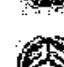 | RT 90° | 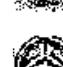 | 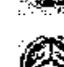 | 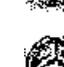 | 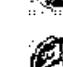 |
| FL Vertical | 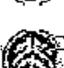 | 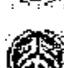 | 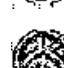 | 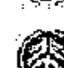 | FL Horizontal | 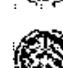 | 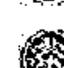 | 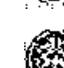 | 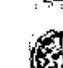 |
| FR 5% | 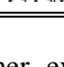 | 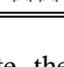 | 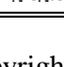 | 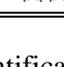 | FD 5% | 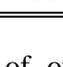 | 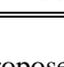 | 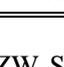 | 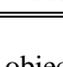 |

To further evaluate the copyright-identification reliability of our proposed RZW-SR$_{3D}$, objective evaluation is then performed. On one hand, we calculate the mean *BERs* between the original watermark and watermarks recovered from different 3D videos when utilizing original 2D frames, utilizing depth maps, and fusing these two components together. The results are compared with that of AUEW and those of RZW$_{3D}$, as shown in Table 6. UVW is not included in the subjective evaluation because the watermarks in the UVW-based ones are only viewable but can be hardly extracted for objective evaluation. On the other hand, we calculate the mean *BERs* when utilizing synthesized frames and fusing them with their depth maps to evaluate the watermark robustness against DIBR distortions. The results are shown in Table 7.

Table 6 shows that the mean *BERs* of our proposed RZW-SR$_{3D}$ when utilizing original 2D frames, utilizing depth maps and fusing them together are very small and their maximal values are 0.1783, 0.1700 and 0.0870, respectively. These values are smaller than the maximal value of AUEW, which is 0.5873, and those of RZW$_{3D}$, which are 0.4981, 0.4942 and 0.4772, respectively. The average values of the mean *BERs* of our proposed RZW-SR$_{3D}$ are merely 0.0619, 0.0628, and 0.0287, respectively. These values are also much smaller than that of AUEW, which is 0.1759, and those of RZW$_{3D}$, which are 0.1308, 0.1706 and 0.1185. Especially, the mean *BERs* of RZW-SR$_{3D}$ are much smaller than those of AUEW under severe filtering, heavy noise addition, rotation, and flipping attacks. Meanwhile, the mean *BERs* of RZW-SR$_{3D}$ are much smaller than those of RZW$_{3D}$ under logo insertion, cropping, rotation, and flipping attacks. These results demonstrate that the copyright-identification of RZW-SR$_{3D}$ is effective under all the tested attacks and much more reliable than those of AUEW and RZW$_{3D}$. As analyzed in the section 3, the sufficient robustness of RZW-SR$_{3D}$ against the severe filtering and heavy noise-addition attacks is due to the utilization of centroids of normalized TIRI-based deviations, and the sufficient robustness of RZW-SR$_{3D}$ against the logo insertion, cropping, rotation and flipping attacks is due to the utilization of concentric-based partition and calculation of deviation between the center pixels and their 8 spatial-neighborhood pixels.

**TABLE 6.** Objective evaluation of copyright-identification reliability in terms of mean *BER* when utilizing original 2D frames, utilizing depth maps and fusing them

|  | AUEW [25] | RZW$_{3D}$ [26] | | | RZW-SR$_{3D}$ | | |
| --- | --- | --- | --- | --- | --- | --- | --- |
|  |  | 2D frame | depth map | Two together | 2D frame | depth map | Two together |
| GB 9×9 | 0.0000 | 0.0195 | 0.0597 | 0.0144 | 0.0435 | 0.0216 | 0.0144 |
| GB 15×15 | 0.0000 | 0.0206 | 0.0613 | 0.0077 | 0.0435 | 0.0216 | 0.0144 |
| AF 9×9 | 0.0007 | 0.0444 | 0.0922 | 0.0384 | 0.0822 | 0.0360 | 0.0250 |
| AF 15×15 | 0.4731 | 0.0663 | 0.1184 | 0.0578 | 0.1075 | 0.0478 | 0.0331 |
| MF 9×9 | 0.0009 | 0.0336 | 0.0728 | 0.0247 | 0.0605 | 0.0172 | 0.0134 |
| MF15×15 | 0.4590 | 0.0573 | 0.1069 | 0.0461 | 0.0818 | 0.0217 | 0.0171 |
| CC -30 | 0.0005 | 0.0144 | 0.0408 | 0.0083 | 0.0210 | 0.0217 | 0.0106 |
| CC +30 | 0.1319 | 0.0277 | 0.0947 | 0.0233 | 0.0319 | 0.0815 | 0.0229 |
| CB -30 | 0.0277 | 0.0264 | 0.0483 | 0.0155 | 0.0698 | 0.0786 | 0.0369 |
| CB +30 | 0.1343 | 0.0239 | 0.1022 | 0.0197 | 0.0595 | 0.1068 | 0.0382 |
| GT $\gamma$ =0.6 | 0.0001 | 0.0492 | 0.0659 | 0.0375 | 0.0463 | 0.0473 | 0.0234 |
| GT $\gamma$ =1.4 | 0.0162 | 0.0414 | 0.0558 | 0.0178 | 0.0367 | 0.0425 | 0.0197 |
| GN 0.005 | 0.1615 | 0.0180 | 0.0422 | 0.0106 | 0.0424 | 0.0922 | 0.0291 |
| GN 0.01 | 0.4184 | 0.0206 | 0.0464 | 0.0134 | 0.0632 | 0.1197 | 0.0413 |
| LI 32×32 | 0.0009 | 0.2075 | 0.2797 | 0.1823 | 0.0000 | 0.0115 | 0.0000 |
| LI 64×64 | 0.0059 | 0.2919 | 0.3459 | 0.2648 | 0.0475 | 0.0451 | 0.0231 |
| RS 1/2 | 0.0000 | 0.0141 | 0.0347 | 0.0067 | 0.0329 | 0.0147 | 0.0101 |
| RS 1/5 | 0.0004 | 0.0139 | 0.0342 | 0.0075 | 0.0643 | 0.0220 | 0.0164 |
| CR 5% | 0.1086 | 0.1578 | 0.3308 | 0.1570 | 0.0459 | 0.0650 | 0.0269 |
| CR 10% | 0.2139 | 0.2330 | 0.3625 | 0.2236 | 0.0782 | 0.1099 | 0.0457 |
| RT 45 | 0.5873 | 0.4567 | 0.4742 | 0.4394 | 0.1422 | 0.1458 | 0.0719 |
| RT 90 | 0.5820 | 0.4898 | 0.4867 | 0.4678 | 0.1783 | 0.1700 | 0.0870 |
| FL Vertical | 0.5826 | 0.4981 | 0.4942 | 0.4772 | 0.0112 | 0.0180 | 0.0069 |
| FL Horizontal | 0.5863 | 0.4906 | 0.4895 | 0.4720 | 0.0105 | 0.0178 | 0.0066 |
| FR 5% | 0.0406 | 0.0325 | 0.0441 | 0.0180 | 0.0943 | 0.1638 | 0.0598 |
| FD 5% | 0.0422 | 0.0519 | 0.0509 | 0.0298 | 0.1153 | 0.0942 | 0.0518 |
| Average value | 0.1759 | 0.1308 | 0.1706 | 0.1185 | 0.0619 | 0.0628 | 0.0287 |

**TABLE 7.** Objective evaluation of copyright-identification reliability in terms of mean *BER* when utilizing depth maps, utilizing synthesized frames, and fusing them

| | Depth map | Baseline distance=5% frame width | | | | Baseline distance=7% frame width | | | |
| --- | --- | --- | --- | --- | --- | --- | --- | --- | --- |
| | | Synthesized frame (left) | Two together | Synthesized frame (right) | Two together | Synthesized frame (left) | Fuse two together | Synthesized frame (right) | Two together |
| Without attack | 0.0000 | 0.0495 | 0.0000 | 0.0492 | 0.0000 | 0.0670 | 0.0000 | 0.0670 | 0.0000 |
| GB 9×9 | 0.0216 | 0.0692 | 0.0114 | 0.0695 | 0.0115 | 0.0838 | 0.0119 | 0.0844 | 0.0120 |
| GB 15×15 | 0.0216 | 0.0693 | 0.0114 | 0.0695 | 0.0115 | 0.0838 | 0.0119 | 0.0844 | 0.0120 |
| AF 9×9 | 0.0360 | 0.0977 | 0.0267 | 0.0975 | 0.0267 | 0.1082 | 0.0272 | 0.1089 | 0.0270 |
| AF 15×15 | 0.0478 | 0.1182 | 0.0385 | 0.1180 | 0.0389 | 0.1265 | 0.0391 | 0.1264 | 0.0391 |
| MF 9×9 | 0.0172 | 0.0799 | 0.0086 | 0.0974 | 0.0085 | 0.0928 | 0.0088 | 0.0939 | 0.0087 |
| MF15×15 | 0.0217 | 0.0950 | 0.0136 | 0.0805 | 0.0136 | 0.1053 | 0.0138 | 0.1073 | 0.0135 |
| CC -30 | 0.0217 | 0.0578 | 0.0111 | 0.0581 | 0.0111 | 0.0744 | 0.0117 | 0.0738 | 0.0116 |
| CC +30 | 0.0815 | 0.0645 | 0.0432 | 0.0639 | 0.0430 | 0.0801 | 0.0494 | 0.0786 | 0.0493 |
| CB -30 | 0.0786 | 0.0914 | 0.0563 | 0.0905 | 0.0560 | 0.1044 | 0.0598 | 0.1029 | 0.0589 |
| CB +30 | 0.1068 | 0.0824 | 0.0590 | 0.0820 | 0.0589 | 0.0944 | 0.0646 | 0.0941 | 0.0643 |
| GT γ =0.6 | 0.0473 | 0.0704 | 0.0305 | 0.0701 | 0.0305 | 0.0846 | 0.0321 | 0.0842 | 0.0326 |
| GT γ =1.4 | 0.0425 | 0.0629 | 0.0264 | 0.0641 | 0.0263 | 0.0782 | 0.0279 | 0.0789 | 0.0277 |
| GN 0.005 | 0.0922 | 0.0676 | 0.0477 | 0.0695 | 0.0492 | 0.0832 | 0.0530 | 0.0839 | 0.0543 |
| GN 0.01 | 0.1197 | 0.0834 | 0.0625 | 0.0847 | 0.0644 | 0.0966 | 0.0686 | 0.0976 | 0.0705 |
| LI 32×32 | 0.0115 | 0.0495 | 0.0017 | 0.0492 | 0.0015 | 0.0670 | 0.0019 | 0.0670 | 0.0018 |
| LI 64×64 | 0.0451 | 0.0772 | 0.0282 | 0.0755 | 0.0278 | 0.0918 | 0.0294 | 0.0895 | 0.0294 |
| RS 1/2 | 0.0147 | 0.0630 | 0.0053 | 0.0626 | 0.0052 | 0.0788 | 0.0055 | 0.0785 | 0.0055 |
| RS 1/5 | 0.0220 | 0.0841 | 0.0135 | 0.0837 | 0.0132 | 0.0970 | 0.0137 | 0.0971 | 0.0137 |
| CR 5% | 0.0650 | 0.0769 | 0.0452 | 0.0765 | 0.0457 | 0.0926 | 0.0485 | 0.0916 | 0.0489 |
| CR 10% | 0.1099 | 0.1017 | 0.0701 | 0.1014 | 0.0700 | 0.1148 | 0.0749 | 0.1144 | 0.0753 |
| RT 45 | 0.1458 | 0.1504 | 0.1056 | 0.1547 | 0.1085 | 0.1568 | 0.1079 | 0.1615 | 0.1104 |
| RT 90 | 0.1700 | 0.1811 | 0.1289 | 0.1860 | 0.1302 | 0.1855 | 0.1306 | 0.1907 | 0.1317 |
| FL Vertical | 0.0180 | 0.0405 | 0.0077 | 0.0584 | 0.0082 | 0.0585 | 0.0081 | 0.0753 | 0.0084 |
| FL Horizontal | 0.0178 | 0.0508 | 0.0078 | 0.0504 | 0.0075 | 0.0679 | 0.0080 | 0.0675 | 0.0078 |
| FR 5% | 0.1638 | 0.1096 | 0.0896 | 0.1088 | 0.0895 | 0.1205 | 0.0960 | 0.1194 | 0.0961 |
| FD 5% | 0.0942 | 0.1292 | 0.0787 | 0.1298 | 0.0798 | 0.1382 | 0.0798 | 0.1397 | 0.0811 |
| Average value | 0.0628 | 0.0842 | 0.0381 | 0.0852 | 0.0384 | 0.0975 | 0.0402 | 0.0985 | 0.0404 |

In addition, Table 6 demonstrates that the mean $BER_S$ when fusing 2D frames and depth maps together are even smaller than those when utilizing only one of them, which demonstrates that attention-based fusion enhances the copyright-identification performance. The reason for this phenomenon is due to utilization of the complementary feature robustnesses of 2D frame and depth map, as analyzed in the evaluation of the performance of similarity-based retrieval.

Table 7 shows that mean *BERs* when utilizing synthesized frames are still very small with their maximal value as 0.1811, 0.1860, 0.1855, 0.1907, and average value as 0.0842, 0.0852, 0.0975, 0.0985. Especially, for the synthesized frames without attacks, their *BERs* are as small as 0.0495, 0.0492, 0.0670 and 0.0670.

These results demonstrate that the performances of copyright-identification in our proposed RZW-SR$_{3D}$ when utilizing synthesized frames are still remarkable although they are slightly worse than those when utilizing original 2D frames. Moreover, when fusing the synthesized frames and depth maps, the maximal values of mean *BERs* values are 0.1289, 0.1302, 0.1306, 0.1317 and average values are 0.0381, 0.0384, 0.0402, 0.0404. These values are even smaller than those when utilizing only the synthesized frames or depth maps, which is consistent with the results listed in Table 6.

## 4.4 Qualitative comparisons with current state-of-the-art schemes

In this experiment, qualitative comparisons with other watermark schemes are conducted in terms of 5 aspects: 1) reliability of the copyright identification, 2) accuracy of the copyright identification, 3) distortion of DIBR 3D video, 4) involvement of similarity-based retrieval, and 5) independence of the copyrights of the 2D frame and depth map. The comparison results are shown in Table 8.

**TABLE 8.** Qualitative comparisons with other schemes

|  | 2D frame-based watermark | Depth map-based watermark | Zero-watermark | RZW-SR$_{3D}$ |
| --- | --- | --- | --- | --- |
| Reliability | Sufficient | **Insufficient** | Sufficient | Sufficient |
| Accuracy | Sufficient | Sufficient | **Insufficient** | Sufficient |
| Distortion of 3D video | **Yes** | No | No | No |
| Similarity-based retrieval | No | No | **No** | Yes |
| Independence of copyrights | **No** | **No** | Yes | Yes |

As shown in Table 8, our proposed RZW-SR$_{3D}$ outperforms other state-of-the-art schemes in terms of all these 5 aspects. In contrast to 2D frame-based watermark schemes [1-3, 6-7, 9, 13, 16, 19, 22, 27, 30, 32, 43], RZW- SR$_{3D}$ does not directly embed watermark information into 2D frames, which avoids introducing distortion into the 3D videos. Compared with the depth map-based watermark schemes [8, 24-25, 29, 33-34, 38], RZW-SR$_{3D}$ is more robust against various video attacks and thus can provide more reliable copyright identification. More importantly, 2D frame-based and depth map-based watermark schemes has rarely considered the possibility that the copyrights of the 2D frame and depth map of synthesized DIBR 3D videos may be different. In this situation, their copyrights should better be protected independently. Only the RZW$_{3D}$ [26] and RZW-SR$_{3D}$ have designed flexible mechanisms for copyright identification and utilized the features of the 2D frame and depth map separately to satisfy this special DRM requirement of DIBR 3D videos. In contrast to the RZW$_{3D}$, RZW-SR$_{3D}$ designs an effective similarity-based retrieval mechanism and their relevant ownership can be obtained precisely according to the retrieval results. Therefore, the accuracy of copyright identification in RZW-SR$_{3D}$ are ensured when processing large-scale videos, which cannot be achieved by RZW$_{3D}$.

## 5. Conclusions

In this paper, a novel RZW-SR$_{3D}$ is proposed by fusing the similarity-based retrieval and zero-watermark technique. It is the first scheme which fuses these two techniques for protecting DIBR 3D videos. Our analytical and experimental results have demonstrated that our proposed RZW-SR$_{3D}$ possesses the following merits. First, RZW-SR$_{3D}$ does not directly embed the copyright information into the 3D videos. Therefore, RZW-SR$_{3D}$ can avoid content distortion, which outperforms 2D frame-based watermark schemes. Second, RZW-SR$_{3D}$ achieves significant robustness against global, geometrical, temporal attacks and DIBR distortions by utilizing temporal information and designing invariant video features, which outperforms depth map-based watermark schemes. Third, RZW-SR$_{3D}$ can retrieve and protect 2D frame and depth map of 3D videos independently and simultaneously by designing flexible mechanisms based on attention-based fusion. It better satisfies different DRM requirements of 3D videos compared to 2D frame and depth map-based schemes. Finally, RZW-SR$_{3D}$ can obtains the ownership relevant to 3D videos precisely and reliably even when processing large-scale DIBR 3D videos by designing an effective similarity-based retrieval mechanism, which cannot be achieved by the zero-watermark scheme for DIBR 3D videos. In the future, we will explore the possibility of applying our proposed scheme to protect medical volume data such as 3D computed tomography (CT) imaging, magnetic resonance imaging (MRI) and optical coherence tomography (OCT) imaging.

## Acknowledgements

This research is supported by the National Nature Science Foundations of China [61602527, 61702558, 61573380, 61772555], and Hunan Provincial Natural Science Foundations of China [2017JJ3416, 2017JJ3411, 2018JJ2548]. The authors would like to thank for the supports from postdoctoral research station of Central South University and the valuable comments on this paper from Prof. Shuo Li, Prof. Qi Tian and Dr. Guibo Luo.